\documentclass{article}
\usepackage[T1]{fontenc}
\usepackage[utf8]{inputenc}
\usepackage[]{ismir} 
\usepackage{amsmath,cite,url}
\usepackage{graphicx}
\usepackage{color}

\usepackage{pifont}
\usepackage[dvipsnames]{xcolor}
\usepackage{bbding}
\definecolor{mred}{RGB}{198, 71, 86}
\usepackage[utf8]{inputenc}
\usepackage{wasysym}
\usepackage{tikz}

\usepackage{fdsymbol}
\usepackage{caption}
\usepackage{subcaption}
\usepackage{graphicx}
\usepackage{booktabs}
\usepackage{multirow}
\usepackage{dirtytalk}
\usepackage{xurl}
\usepackage{enumitem}

\title{MusGO: A Community-Driven Framework \\for Assessing Openness in Music-Generative AI}

\multauthor
  {Roser Batlle-Roca$^1$ \hspace{1cm} Laura Ibáñez-Martínez$^1$ \hspace{1cm} Xavier Serra$^1$}
  {{\bf Emilia Gómez$^{1,2}$ \hspace{1cm} Martín Rocamora$^1$ \hspace{1cm}}\\
  $^1$ Music Technology Group, Universitat Pompeu Fabra, Barcelona Spain\\
  $^2$ Joint Research Centre, European Commission, Seville, Spain\\
  {\tt\small \{roser.batlle, laura.ibanez, martin.rocamora\}@upf.edu}
  }

\def\authorname{R. Batlle-Roca, L. Ibáñez-Martínez, X. Serra, E. Gómez, and M. Rocamora}

\usepackage[bookmarks=false,pdfauthor={\authorname},pdfsubject={\pdfsubject},hidelinks]{hyperref}

\begin{document}

\maketitle

\begin{abstract}
Since 2023, generative AI has rapidly advanced in the music domain. Despite significant technological advancements, music-generative models raise critical ethical challenges, including a lack of transparency and accountability, along with risks such as the replication of artists’ works, which highlights the importance of fostering openness.
With upcoming regulations such as the EU AI Act encouraging open models, many generative models are being released labelled as ‘open’. However, the definition of an open model remains widely debated. In this article, we adapt a recently proposed evidence-based framework for assessing openness in LLMs to the music domain. 
Using feedback from a survey of 110 participants from the Music Information Retrieval (MIR) community, we refine the framework into MusGO (Music-Generative Open AI), which comprises 13 openness categories: 8 \textit{essential} and 5 \textit{desirable}.
We evaluate 16 state-of-the-art generative models and provide an openness leaderboard that is fully open to public scrutiny and community 
contributions. Through this work, we aim to clarify the concept of openness in music-generative AI and promote its transparent and responsible development.
\end{abstract}

\vspace{-0.3cm}
\section{Introduction}\label{sec:introduction}

Music-generative AI is introducing 
critical ethical concerns, particularly regarding its impact on creative processes and authorship, potential legal issues from data misuse, and disruptions to existing business and intellectual property (IP) models~\cite{Carnovalini2020, Gomez2018, Strum2019, Barnett2023, BatlleRoca2024}. Furthermore, these technologies usually exhibit a Western 
cultural bias, undermining diversity in musical expression 
and homogenising music genres.
Additionally, they often 
require substantial computational resources, 
contributing to energy consumption amid the climate crisis~\cite{Barnett2023, Tatar2024}.
A further concern is the lack of transparency in these models, 
which limits scrutiny of their behaviour, attribution, and decision-making \cite{BatlleRoca2024},
as well as the potential replication of training data \cite{Carlini2023a, Carlini2023b, bralios2024}.

To address these challenges, we advocate for the use of open models. Open models have a vital role in open science, promoting transparency, accountability, and innovation~\cite{Gundersen2018, liesenfeld2024, spirling2023}. 
Alongside academic efforts \cite{groeneveld2024}, tech companies are increasingly 
labelling AI systems as open. 
However, several initiatives provide only a few components, such as model weights and inference code \cite{liesenfeld2024}, which 
limits reuse and transparency, and leads to criticism of `open-washing'. 
Yet, defining openness for AI models is challenging and remains a subject of active debate.
In fact, open model status is becoming highly attractive with the legal exceptions in the EU AI Act, 
making it timely to examine the requirements needed to attain it \cite{liesenfeld2024, Solaiman2023}.

The absence of a defined methodology for evaluating openness in music-generative models creates a potential gap in their assessment. 
Consequently, we pose the question: 
\textbf{how can openness be assessed in music-generative AI?} 
We draw upon and adapt the work by Liesenfeld and Dingemanse (2024) \cite{liesenfeld2024}, which introduces an evidence-based method for evaluating openness in large language models (LLMs).
Their proposed framework is composite and graded, 
consisting of multiple categories, each with three levels of openness.
Building on and tailoring this approach for the music domain,
we introduce \textbf{MusGO} (Music-Generative Open AI): a community-driven framework for assessing openness in music-generative AI.
We develop MusGO by actively involving the Music Information Retrieval (MIR) community through a survey and 
refining the criteria based on their feedback.

Our main contributions include (1) a reproducible \textbf{framework}\footnote{\tiny{\url{https://roserbatlleroca.github.io/MusGO_framework/framework.html}}}
for evaluating openness in music-generative AI, (2) a \textbf{leaderboard}\footnote{\tiny{\url{https://roserbatlleroca.github.io/MusGO_framework/index.html}}}
showcasing openness across 16 models, and (3) a fully open \textbf{repository}\footnote{\tiny{\url{https://github.com/roserbatlleroca/MusGO_framework}}} to enable public scrutiny, community contribution and future updates. With this, we aim to support future work by providing a more nuanced understanding of openness in music-generative AI, offering a fine-grained evidence-based approach to inform decisions and promote responsible practices.

\section{Background and related work}
\label{sec:background}
\subsection{Defining  ‘open’ models}

Documentation efforts in AI have supported model transparency by disclosing development processes, data sources, and model attributes \cite{Arnold2019, Gebru2021, Holland2018, Mitchell2019}. However, defining openness in AI is challenging, as it involves multiple components (e.g., source code, documentation, model weights, training data)~\cite{groeneveld2024} and requires balancing openness ideals
with practical concerns such as data privacy, security, and IP rights~\cite{desai2024}.
Recent initiatives have contributed to shaping such a definition. 
One of these is the Open Source Initiative (OSI), which has led a public consultation to adapt its widely accepted definition of open-source software to AI systems, resulting in the release of the Open Source AI Definition (OSAID) 1.0 in October 2024~\cite{OSAID2024}.  
It requires an AI system to provide access to its source code and model weights, along with sufficient information about the training data. The goal is to ensure these models
can be freely used, modified, and shared, thereby promoting transparency and collaboration within the AI community.

While OSAID represents an important step toward setting a standard for AI openness, it has also sparked discussion and criticism \cite{Robison2024, Gall2024, openfuture2024}.
Some argue that the stringent requirements may be challenging to meet, 
possibly excluding models that, 
although not fully compliant, still contribute 
meaningfully to the open-source ecosystem.
Another concern raised by open data advocates is that OSAID does not require the training data itself to be shared, only detailed information about it, which is sometimes seen as a form of ‘open-washing’. This is particularly relevant in music, where much of the data, such as recordings, compositions, and lyrics, is owned by rights holders and often cannot be made publicly available. 
The freedom to use a model for any purpose also raises concerns about unintentional copyright violations or the deliberate imitation of an artist’s style without their consent~\cite{Tatar2024, jiang2023, BatlleRoca2024}. In contrast, Responsible AI Licenses (RAIL) propose a more nuanced form of openness by including restrictions on certain uses.

Among other regulatory efforts, the European Union's AI Act~\cite{EUAIAct2024} has granted some exemptions 
to models released under a free and open-source license, as specified in Article 53 (1)(a) and (b) \cite{EUAIAct_Art53}.
The public availability requirement appears to cover \say{\textit{(...) its parameters, including the weights, the information on the model architecture, and the information on model usage}, 
with no explicit mention of the training data~\cite{FAQAI}. 
The \textit{General-Purpose AI Code of Practice} 
is expected to provide further details on the obligations 
under Articles 53 and 55 for different ways of releasing general-purpose AI models, including open sourcing, although the document is still
in draft form~\cite{CodeOfPracticeDraft}. 

Rather than conceiving openness as a binary status, other proposals are tiered, with openness 
distributed across different dimensions~\cite{Solaiman2023, liesenfeld2023}. 
This aligns with prior work on AI accountability and transparency frameworks~\cite{konigstorfer2022, matthews2020, walmsley2021, novelli2024}. In this 
view, a 
model's openness is 
determined by the 
relative openness of its 
individual components, 
accommodating cases 
where} not all 
criteria are fully attained. 

\subsection{How to assess openness?}

Building on the idea that openness is not a binary attribute but rather a composite of various elements, each 
varying across different degrees, several initiatives have aimed to create structured, tiered, gradient-based frameworks to evaluate it~\cite{bommasani2024,white2024,liesenfeld2024, eiras2024b}. 

For instance, the Linux Foundation's Model Openness Framework (MOF)~\cite{white2024} distinguishes three tiers of open AI systems. It promotes completeness and openness for reproducibility, transparency, and usability, categorising AI models 
based on their level of openness and ensuring the release of key artefacts like model architecture, training data, and evaluation results under open licenses. Similarly, the Foundation Model Transparency Index (FMTI)~\cite{bommasani2024} evaluates transparency in foundation models using 100 indicators across three broad domains—upstream, model, and downstream—focusing on aspects such as data, compute resources, model capabilities, and downstream impact. While comprehensive, the FMTI does not allow individual data points to be scrutinised or contested, limiting the transparency and verifiability of its scores~\cite{liesenfeld2024}. 

On the other hand, Eiras et al. \cite{eiras2024a, eiras2024b} have contributed valuable insights into the risks and opportunities associated with open-source generative AI models, proposing a taxonomy for evaluating 
their level of openness, particularly focusing on the availability of code and data. 
Although advocating for AI democratisation, they emphasise the need for openness to be balanced with responsible AI practices to avoid risks such as misuse, bias, and unintended consequences. Their evaluation framework highlights both technical openness (e.g., code and model weights availability) and ethical concerns (e.g., data privacy and model safety), offering a more comprehensive perspective on open-source AI 
compared to other approaches.

A particularly relevant framework is the one
proposed by 
Liesenfeld and Dingemanse (2024) \cite{liesenfeld2024} which, 
building on previous work \cite{liesenfeld2023}, offers an evidence-based method for evaluating openness in LLMs, 
and 
extends it to text-to-image models. The proposed framework is \textbf{composite}—it considers multiple dimensions when determining openness—and \textbf{graded}—meaning each category may have different levels of openness: \textit{closed}, \textit{partial} and \textit{fully open}. Many of these dimensions 
align with those from other frameworks, particularly MOF. The approach has been tested since July 2023 in an open leaderboard\footnote{\tiny{\url{https://opening-up-chatgpt.github.io/}}} that tracks the degree of openness of several LLMs.
A distinguishing feature of this method is its 
community-driven approach, which allows for the inspection and appeal 
of the model's assessment scores and their supporting evidence. 
However, while 
the framework is adaptable to various domains, it is essential to 
revise certain categories and include domain-specific elements to accurately assess openness \cite{liesenfeld2024}. 
In the case of music, unique challenges, such as the copyrighted nature of most training data, underscore the need for tailored criteria and domain-specific practices to ensure a comprehensive openness assessment.

\section{Openness Framework for Music AI}
\label{sec:builingmusgo}

\subsection{From LLMs to music}
\label{sec:llmstomusic}

We adopted the evidence-based framework introduced by Liesenfeld and Dingemanse (2024) \cite{liesenfeld2024}  
and tailored it to the music domain. 
Our initial adaptation involved modifying references to LLMs to align with music-generative models, 
for example, by renaming LLM-oriented labels and excluding instruction tuning-related categories,
which led to a revised framework 
of 11 categories from the original 14. 
To ensure the adapted framework reflected the perspectives and priorities 
of the MIR community, we aimed to gather feedback on each category's intention and relevance, considering the risks of over-reliance on single features, such as access or licensing, to determine openness \cite{liesenfeld2024}. 
To this end, we conducted an anonymous online questionnaire within the MIR community.

\subsection{MIR community survey}
\label{subsec:survey}

\subsubsection{Methodology}

The survey included a statement 
for each category reflecting the top open level (\textit{fully open}). Participants were asked to assess the proposed statement by (1) \textbf{relevance}---how important they consider the category in determining the openness of a model---
using a 5-point Likert scale (1 = \textit{not relevant}, 5 = \textit{very relevant}),
and (2) \textbf{agreement}---whether they agreed, somewhat agreed, or 
disagreed with the proposed statement. They were encouraged to suggest modifications and refinements 
to the statements when they disagreed or had reservations about the criteria. 
The survey also included an open comments section for participants to propose new categories and raise unaddressed concerns. 

The survey was open for 8 weeks and received 
110 responses from participants linked to the MIR community.
Although we intended broader representation, the sample was biased towards male academics in Europe and North America. However, this distribution aligns with the typical
demographics of attendees at
the International Society for Music Information Retrieval (ISMIR) 
conference.\footnote{Complementary information on the survey, including a
detailed analysis of the participants' backgrounds and their responses, is available at:
\tiny{\url{https://roserbatlleroca.github.io/MusGO_framework/survey.html}}}

\subsubsection{Results}
Table \ref{tab:results} summarises the 
survey results, showing the median relevance score ($M$) and agreement levels for each category.
Overall, participants mostly agreed with the proposed statements,
indicating a general consensus on the proposed criteria.
Categories \textit{3: Model Weights}, \textit{4: Code documentation} and \textit{1: Open Code} showed the highest levels of agreement (with over 78\% selecting \textit{Yes, I agree}), and were also rated highly relevant ($M_{1,3}$=5, $M_{4}$=4).
For category \textit{2: Training data}, 
while relevance was high ($M_{2}$=5), there 
was significant discussion regarding the framing of the statement (30\% 
selected \textit{Yes, I somewhat agree}).
These results suggest that elements related to 
code access and model reproducibility are of great interest to the community.

In contrast, categories such as \textit{9: Package} and \textit{10: API}, 
showed lower and more varied agreement rates (46.36\% and 39.09\%, respectively) and lower relevance ($M_{9,10}$=3). 
This was reflected in the comments, which highlighted concerns about the additional effort required to provide these features and questioned their necessity for determining openness.

The survey also highlighted some categories with moderate agreement, such as\textit{ 6: Paper and Preprint}, \textit{7: Model Card} and \textit{8: Datasheet}, 
all with a median relevance score of 4, but with notable disagreement rates. This diversity of responses underscores the varying priorities among community members regarding detailed documentation 
across different areas of the model. 
Comments on these categories were key to understanding the community’s perspectives and refining the initial criteria.

\begin{table}[t]
\centering
\caption{Results from the MIR community survey,  
showing for each category its median relevance and level of agreement (\%) with the proposed statement. Note that the criteria correspond to the first adapted version
(see \ref{sec:llmstomusic}).}
\label{tab:results}
\resizebox{\linewidth}{!}{%
\begin{tabular}{lcccc} 
\toprule
\multicolumn{1}{c}{\multirow{2}{*}{\textbf{Criteria}}} & \multicolumn{1}{c}{\textbf{Relevance}} & \multicolumn{3}{c}{\textbf{Agreement (\%)}}                                                                                                                                                                                                             \\

\cmidrule(l){3-5}
\multicolumn{1}{c}{}                                   & \textit{Median}       
& \begin{tabular}[c]{@{}c@{}}\textit{Yes, I }\\\textit{agree.}\end{tabular} & \begin{tabular}[c]{@{}c@{}}\textit{Yes, I some-}\\\textit{what agree.}\end{tabular} & \begin{tabular}[c]{@{}c@{}}\textit{No, I do }\\\textit{not agree.}\end{tabular}  \\ 
\toprule
1: Open Code                                             & 5                   & \textbf{78.18}                                                          & 20.00                                                                             & 1.82                                                                           \\
2: Training Data                                               & 5                  & \textbf{67.27}                                                          & 30.00                                                                             & 2.73                                                                           \\
3: Model Weights                                            & 5                  & \textbf{79.09}                                                          & 18.18                                                                             & 2.73                                                                           \\
4: Code documentation                                  & 4                   & \textbf{79.09}                                                          & 18.18                                                                             & 2.73                                                                           \\
5: Architecture                                                & 4                  & \textbf{62.73}                                                          & 30.00                                                                             & 7.27                                                                           \\
6: Paper and Preprint                                    & 4                  & \textbf{64.55}                                                          & 21.82                                                                             & 13.64                                                                          \\
7: Model Card                                          & 4                   & \textbf{54.55}                                                          & 30.00                                                                             & 15.45                                                                          \\
8: Datasheet                                                 & 4                   & \textbf{64.55}                                                          & 28.18                                                                             & 7.27                                                                           \\
9: Package                                           & 3                   & \textbf{46.36}                                                          & 32.73                                                                             & 20.91                                                                          \\
10: API                                                        & 3                 & \textbf{39.09}                                                          & 30.00                                                                             & 30.91                                                                          \\
11: Licensing                                             & 4               & \textbf{62.73}                                                          & 28.18                                                                             & 9.09                                                                           \\
\bottomrule
\end{tabular}
}
\vspace{-0.5cm}
\end{table}

\subsection{Refining criteria based on community feedback}

Participants' feedback revealed gaps and disagreements that we addressed by carefully reviewing all comments, identifying recurring issues, and refining the criteria accordingly. The most substantial changes 
involved clarifying the descriptions of each category
and  
defining what constituted 
\textit{partial} and \textit{fully open} for each 
case. We integrated 
music-specific nuances 
raised in the feedback, 
including the challenges of 
sharing IP-protected training data, the need to 
provide sonified examples,  
and the user experience needs of musicians, who 
may have limited technical  
skills and domain-specific software preferences, extending beyond the use of APIs. 
In addition, we held internal discussions with 
several researchers from the Music Technology Group (MTG), which provided further insights that helped 
refine the framework.

Following the survey results regarding category relevance, 
we introduced two distinct levels within our framework: \textbf{essential}, 
for categories considered key to assessing openness, and \textbf{desirable}, for 
those that enhance openness and provide additional value but are not indispensable.
Furthermore, feedback highlighted the absence of a category 
addressing model evaluation procedures. In response, we introduced a new 
category, \textit{Evaluation procedure}, 
covering evaluation data, metrics, 
and results on model performance.
We also 
added the category \textit{Supplementary material page} to 
acknowledge the importance of dedicated websites that provide usage instructions, model demonstrations and sonified outputs, which are essential in music research, where audio cannot be embedded 
within the paper itself.
Additionally, we renamed some categories to better 
reflect their scope: \textit{Open code} $\rightarrow$ \textit{Source code}, \textit{Architecture} $\rightarrow$ \textit{Training procedure},  \textit{Paper} $\rightarrow$ \textit{Research Paper}, and \textit{API} $\rightarrow$ \textit{User-oriented application}. The renamed category \textit{User-oriented application} 
also includes other user-experience approaches, such as real-time applications, 
recognising the need for accessible tools for musicians and composers.
These music-specific adaptations were critical to make the framework relevant for music-generative models, which differ significantly from LLMs.

\subsection{The MusGO framework}

As a result of this modification and iteration process, we built the Music-Generative Open AI 
framework.  
MusGO comprises 13 categories: 8 essential and 5 desirable. 
The essential categories include: (1) \textit{Source code}, (2) \textit{Training data}, (3) \textit{Model weights}, (4) \textit{Code documentation}, (5) \textit{Training procedure}, (6) \textit{Evaluation procedure}, (7) \textit{Research paper}, and (8) \textit{Licensing}, while desirable 
categories are defined by (9) \textit{Model card}, (10) \textit{Datasheet}, (11) \textit{Package}, (12) \textit{User-oriented application}, and (13) \textit{Supplementary material page}.
Essential categories follow an openness-graded scale of 3 levels: 
closed (\ding{55}), partial (\textbf{$\sim$}) and fully (open) (\ding{51}). 
Instead, desirable
categories are binary: whether that element exists (\textcolor{Goldenrod}{$\medblackstar$}) or not, 
as they are considered complementary add-ons to the model, rather than core components, which allows 
supporting any efforts made in these areas. 
A complete description of the criteria is available on our complementary website.

\vspace{-0.25cm}

\section{Assessing Openness}
\label{sec:open}
\subsection{Model selection}

We selected 16 state-of-the-art 
music generation models:\footnote{Introduced from oldest to newest research paper release.} 
GANSynth \cite{engel2018}, Jukebox \cite{dhariwal2020}, RAVE \cite{caillon2021}, Musika \cite{pasini2022}, Moûsai \cite{schneider2023}, MusicGen \cite{Copet2024}, MusicLM \cite{agostinelli2023}, VampNet \cite{garcia2023}, MusicLDM \cite{Chen2024}, Music ControlNet \cite{wu2024},
Noise2Music \cite{huang2023}, MeLoDy \cite{lam2024}, DITTO-2 \cite{Novack2024}, Diff-A-Riff \cite{nistal2024}, JASCO \cite{tal2024}, and Stable Audio Open \cite{evans2024open}. 
Our selection encompasses a 
diverse set of music-generative models with waveform outputs,
mainly focusing on those that explicitly claim to be ‘open’ or are associated with openness, 
alongside popular models that exhibit openness behaviours (e.g., publishing model weights, releasing research papers, or providing 
user-oriented applications). In addition to this, we aimed to cover a 
variety of architectures commonly used within the field, including
Variational Autoencoders (VAEs) \cite{kingma2014}, Generative Adversarial Networks (GANs) \cite{goodfellow2020}, Transformers \cite{vaswani2017}, Diffusion models \cite{ho2020}, and novel methodologies such as Flow Matching \cite{lipman2023}.

 \vspace{-0.2cm}
\subsection{MusGO into practice}

We evaluated the selected models using a structured 
methodology based on a checklist of fully open level requirements
for each category. 
Each model was examined by one of the authors, including justification statements.
Two other authors then reviewed the evaluations to  ensure consistency and minimise bias. 
In case of discrepancy, we engaged in 
discussions and adjusted the assessment iteratively, 
following consensual qualitative research principles \cite{hill1997}.
We gathered information from official publishers and maintainers and, when available, complemented 
it with third-party sources such as HuggingFace.\footnote{\tiny{\url{https://huggingface.co/}}}

\vspace{-0.15cm}

\subsection{Openness assessment}

\begin{figure*}[ht]
    \centering
    \includegraphics[width=\textwidth]{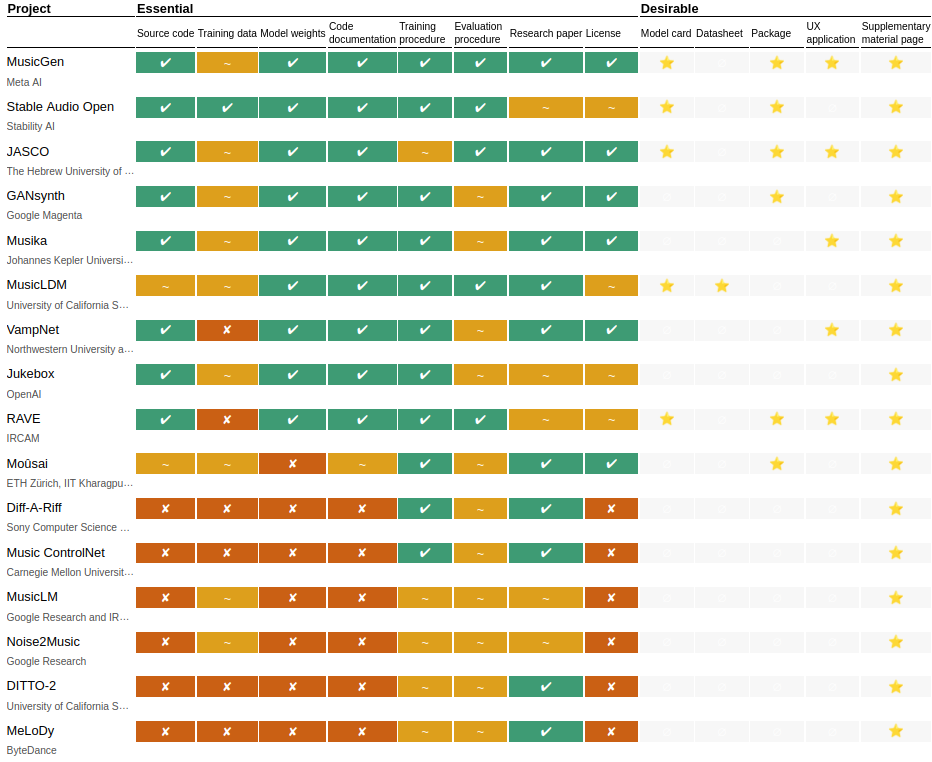}
    \caption{Openness leaderboard 
    of 16 music-generative models assessed
    using the MusGO framework.}
    \label{fig:evaluation}
\end{figure*}

Figure \ref{fig:evaluation} 
presents the openness evidence-based assessment of the 16 selected models, 
forming the basis of our openness leaderboard.
Models are 
ordered using a weighted openness score ($O$), based on essential categories ($E$) and normalised to a 100-point scale. 
Considering survey findings, the three most relevant categories ($E_{1}$, $E_{2}$ and $E_{3}$, all with $M_{1,2,3}$=5) are weighted twice as much as the others. 
Note that the score 
is used for ordering purposes only, and we do not intend to reduce openness to a single value. When models achieve the same score, the order 
is determined by the highest number of fulfilled desirable
categories. 

The assessment reveals significant variation in openness across the considered models.
The category \textit{Training procedure} is the most open,
with 11 models  
fully open and 5 partially open. Instead, the category \textit{Training data} is the most closed, 
with only one model (Stable Audio Open \cite{evans2024open}) achieving fully open status.
Regarding source code, 
eight models provide full access, 
and two provide partial access. The 
remaining six models, 
besides not providing source code, also 
do not disclose model weights, code documentation, 
or apply appropriate licensing. Models classified as fully open for category \textit{Model weights} also 
tend to provide comprehensive documentation of their code and 
are typically licensed under an OSI or RAIL license,
reflecting a correlation between source code, model weights, code documentation, and licensing. Regarding the category \textit{Evaluation procedure}, most models (11/16) are 
partially open, often lacking evaluation data or implementation details, while only five are fully open. 
Although all models analysed
provide a research paper or equivalent technical report, only 11 meet the peer-reviewed and accessibility requirements. 
Desirable categories also show a substantial variability across models. The category \textit{Datasheet} is the least fulfilled,
with only one model (MusicLDM \cite{Chen2024}) 
providing detailed and structured data documentation. Instead, all models 
include a supplementary material page, which demonstrates that accompanying resources with model demonstrations and sound examples
has become a community norm. 
These results are made publicly available through a continuously updated leaderboard,
helping to identify incomplete openness claims and potential cases of `open-washing'. 

\section{Discussion}

\label{sec:discussion}
\subsection{Music-generative open AI}

Our assessment reveals a significantly diverse landscape of music-generative models in terms of openness, highlighting both notable efforts and significant room for improvement.
This situation underscores different 
levels of commitment across the community, particularly regarding critical categories such as training data, source code, 
and licensing. Although the analysed models generally perform well in providing documentation, 
the level of detail is not always sufficient to guarantee complete transparency and accountability. 
Furthermore, desirable  
categories are less commonly 
addressed, especially datasheets, 
pointing to a gap in the  
information available about these models.

Our evaluation validates the MusGO framework as a comprehensive evidence-based resource for assessing openness in music-generative AI. Using a composite and graded framework allows for a more flexible and fine-grained perspective on model openness. For instance, while \textit{Training procedure} is often the most open category, \textit{Training data} remains critical, with the majority of models lacking sufficiently detailed descriptions of the data used. However, we acknowledge that sharing training data in the music domain may pose significant challenges due to IP rights, which can lead to ethical and legal concerns when publishing this information. We address this issue by considering \textit{Training data} fully open when direct access is restricted due to legal concerns, provided that detailed information about all sources is disclosed.

We present a continuously updated leaderboard showcasing openness assessments using the MusGO framework. It reflects the current state of openness in music-generative models and acts as a live collaborative platform for the community. Researchers, developers, and stakeholders are invited to actively contribute by evaluating new models and engaging in discussion. This approach enables tracking model evolution and integrating emerging aspects like controllability and real-time use. The repository provides guidelines on how to contribute, including the detailed framework criteria, instructions for issues and pull requests, and information on submissions review.\footnote{\tiny{\url{https://github.com/roserbatlleroca/MusGO_framework/tree/main/projects/README.md}}}

\subsection{Beyond openness}

While openness is crucial for ensuring transparency and accountability in music-generative models, it alone does not guarantee ethical behaviour or address broader implications. Several survey participants also raised this concern. Barnett \cite{Barnett2023} demonstrates that most literature on generative AI in music tends to overlook the potential negative ethical consequences of these models. However, our analysis suggests a shift with increasing attention focused on these concerns. Among the analysed models, 62.5\% (10/16) include ethical considerations, with a notable rise in papers published after 2023 (83.3\% of these models).

Openness can help reveal where ethical risks arise, especially in sensitive areas such as copyright. While sharing code and data supports transparency, it does not address the legality of the data used for training. For instance, MusicGen \cite{Copet2024} and JASCO \cite{tal2024} obtained data through proper legal agreements, whereas Stable Audio Open \cite{evans2024open} is the only model using fully available and accessible data. This highlights the need to balance openness with careful copyright considerations. Thus, MusGO helps identify and evaluate whether ethical claims are substantiated.

Similarly, openness can point out whether efforts to mitigate harmful or inappropriate uses have been implemented. Models like MusicLDM \cite{Chen2024} and Stable Audio Open \cite{evans2024open} try to mitigate risks by performing memorisation analysis to help ensure that training data is not inadvertently reproduced by their models. In addition, MusicLM \cite{agostinelli2023} and Noise2Music \cite{huang2023} address the risks of plagiarism and cultural appropriation by incorporating safeguards to mitigate such issues. While MusGO does not directly evaluate these aspects, future versions of the framework could help emphasise risk mitigation strategies.

Another key concern is the representation and diversity of the training data. Open datasets may still be biased, with limited cultural diversity. For example, JASCO \cite{tal2024} acknowledges that its dataset is heavily Western-centric, which limits the diversity of the generated music. Similarly, Noise2Music \cite{huang2023} points out that biases in training data can misrepresent musical genres. These examples show that MusGO can help expose limitations in cultural diversity and representation by making dataset composition and documentation more accessible for scrutiny.

Finally, the economic impact of generative AI on musicians and creative professionals is also a critical concern. Open models increase accessibility, but they can potentially displace human creators. Many authors stress that generative models should enhance, not replace, human creativity. Yet, concerns remain about the disruption of traditional industries and the economic impact on musicians. This emphasises the need for openness to be paired with a broader discussion about the ethical, social, and economic implications of AI, including potential usage restrictions. Thus, ethical guidelines and safeguards are essential to ensure that AI-generated music does not perpetuate harmful stereotypes or infringe on the rights of original creators. MusGO offers a structured lens to review models' governance and serve as a foundation for ethical scrutiny.

\subsection{Limitations}

Evaluating the category \textit{Training data} has proven particularly challenging. In the music domain, data accessibility is often restricted by IP rights and legal constraints, and even revealing training data might require a closed audit process to comply with these requirements. Incorporating such a level of detail into our framework would require a more complex approach, which may not be feasible in this context. Moreover, while hardware requirements are considered within the \textit{Training procedure} category, their impact on openness has not been thoroughly analysed. Some models can be trained on personal computers, while others require extensive computational resources, complicating reproducibility and limiting accessibility. For models with lower resource demands, categories like \textit{Training data} may become less relevant, shifting the focus toward enabling artists to train models on their own music.

In developing the MusGO framework, we surveyed the MIR community. However, we acknowledge that our sample was biased towards male academics based in Europe and North America. Expanding the sample in the future to include artists and other stakeholders, as well as perspectives from a broader range of regions, could provide valuable insights and lead to adaptations that would make the framework more inclusive.

While MusGO provides a robust assessment of openness, the leaderboard does not reflect the broader ethical and societal implications of generative AI. This remains a key direction for future improvement, especially as implications related to hardware requirements, real-time use, and accessibility are highly relevant to artists.

\vspace{-0.3cm}
\section{Conclusion}
\label{sec:conclusion}

With the rise of music-generative AI, debates around the ethical implications of these models have intensified. Given the lack of transparency and accountability in these systems, we advocate for open models. Yet, what constitutes an open model remains undefined for music-generative AI. In this work, we adapt an existing openness framework for LLMs to music-generative AI. A community-specific survey helped us identify gaps and particular considerations unique for the music domain. As a result of tailoring the criteria based on the gathered feedback, we introduce the MusGO framework. We put MusGO into practice by analysing 16 state-of-the-art models, providing a public and updatable openness leaderboard. Our analysis reveals that openness in music-generative AI is a field in progress, where significant gaps remain in the availability of training data, model weights, and licensing. MusGO allows for a structured and flexible perspective on model openness by considering complementary characteristics. It reports on a model's status and highlights potential cases of `open-washing'. As generative models continue to evolve, we strive for more consistent and thorough openness practices, particularly when sharing training data and source code, to foster transparency, accountability, and responsible development in music-generative AI.

\section{Ethics Statement}

This study involved a voluntary, anonymous online survey aimed at gathering feedback from the MIR community on a preliminary adapted openness framework. Participants were informed about the scope and purpose of the study, as well as the intended use of the collected data. Regarding study design, participant information and data protection, we adhered to the guidelines and recommendations of the Institutional Committee for Ethical Review of Projects (CIREP) at Universitat Pompeu Fabra. In line with the General Data Protection Regulation (GDPR) 2016/679 (EU), all responses were anonymised and stored securely. Participants were informed of their data rights, including access, rectification, deletion, and withdrawal of consent, in accordance with GDPR protocols.

\section{Acknowledgments}

This work has been supported by \textit{IA y Música: Cátedra en Inteligencia Artificial y Música} (TSI-100929-2023-1), funded by the Secretaría de Estado de Digitalización e Inteligencia Artificial and the European Union-Next Generation EU, and \textit{IMPA: Multimodal AI for Audio Processing} (PID2023-152250OB-I00), funded by the Ministry of Science, Innovation and Universities of the Spanish Government, the Agencia Estatal de Investigación (AEI) and co-financed by the European Union. We thank our colleagues at the Music Technology Group at Universitat Pompeu Fabra for their thoughtful insights, constructive discussions and active engagement throughout the development of this work.

\bibliography{ISMIR}

\end{document}